# When More Is Less: Higher Magnetic Fields and Their Limited Impact on SNR per Time Unit of Acquisition Time in Single Voxel Spectroscopy

Guodong Weng[1,2,3], and Johannes Slotboom[1,2]

Affiliations:
1. Institute for Diagnostic and Interventional Neuroradiology, Inselspital, University Hospital and University of Bern, Switzerland
2. Translational Imaging Center, sitem-insel, Bern, Switzerland
3. Yale School of Medicine, Department of Radiology & Biomedical Imaging, New Haven, CT, USA.

Johannes Slotboom: johannes.slotboom@insel.ch

Guodong Weng: guodong.weng@unibe.ch

# Abstract

Magnetic resonance spectroscopy (MRS) offers significant diagnostic potential but is inherently constrained by low signal-to-noise ratio (SNR). While increasing the main magnetic field strength $B_0$ is theoretically linked to increased SNR, practically obtained gains in SNR from $B_0^{7/4}$ to $B_0$ depending on the domination of thermal noise at high $B_0$, are not always realized. Especially in clinical settings the maximum reachable SNR is further constrained by the total available acquisition time (TA) and the regulatory limits on maximum tolerable specific absorption rate (SAR).

This work attempts to derive mathematical expressions that enable systematical analysis of the theoretical achievable SNR-gain. One important notion is this context is the *SNR gain per unit of measurement time* as a function of the main magnetic field $B_0$ strengths in the case of single voxel spectroscopy (SVS) pulse sequences. Our findings indicate that under given fixed total amount of (patient acceptable) measurement time TA, and maximum tolerable SAR limitation, together with conditions that ensure the adiabaticity of specific sequences, further increasing $B_0$ does not further improve SNR per unit of measurement time. Key factors were identified, including RF-pulse bandwidth scaling with $B_0$ and longitudinal relaxation time ($T_1$) dependencies, impact the net gain as well. Our theoretical analysis emphasizes critical considerations for optimizing *SNR per unit time* in clinical MRS, even challenging the presumption that higher magnetic fields $B_0$ always yield improved SNR per time unit of measurement time performance.

# Introduction

Magnetic Resonance Spectroscopy (MRS) is inherently characterized by low signal-to-noise ratio (SNR), even for protons, which possess the highest gyromagnetic ratio *and* natural abundance. Since the SNR of the MR-signal increases with the main magnetic field strength $B_0$, MRS should ideally be performed at the highest possible magnetic field strength; and marketing of MR-scanners with higher $B_0$ follow this simplistic assumption. The following relation between the magnitude of the equilibrium magnetization $M_0$ and the main magnetic field $B_0$ exists [1]:

$$M_0 = N \gamma^2 \hbar^2 I(I+1) B_0 / 3 k_B T_s \qquad (1)$$

in which $N$ is the number of spins, $\gamma$ the gyro magnetic ratio, $\hbar$ the reduced Planck constant, $I$ the spin quantum number $I$ ($I=½$ for protons), $k_B$ is Boltzmann constant, and $T_s$ is the temperature of the spin system.

In case the equilibrium magnetization $M_0$ is excited by a 90-degree by an RF pulse, and flipped into the transverse plane ($M_{xy} = M_0 \cdot \sin(\frac{\pi}{2})$), it will induce a voltage in the receiver coil, proportional to the transverse magnetization $M_{xy}$, given by:

$$S_{NMR} \propto d\boldsymbol{M}_{xy}/dt \propto \omega M_{xy} \qquad (2)$$

Since in (2) $\omega = \gamma B_0$, this implies that the NMR signal strength scales with $B_0^2$.

This theoretical relationship has led MR-scanner manufacturers to promote mid-field (MF) and ultra-high field (UHF) scanners based on their ability to achieve higher SNR. However, in practice, the relationship appears to be more nuanced. Previous experimental work has demonstrated the close to ideal SNR gain (60-100%) for MRI and MRS with long TR (5000 ms) from 4T to 7T [2], [3]. In contrast, other studies using SVS/CSI MRS, with shorter TR (1500-3000 ms), have shown less SNR gain (23-73%) from 1.5T to 3T, provided that the same TR and TA are compared, as was shown previously in reports [4], [5], [6], [7]. In addition, there are a few studies reporting an SNR gain of 36-60%, with TR of 3000-7000 ms, from 3T to 7T [8], [9], [10]. These findings may give the impression that higher magnetic field strength *always* results in increased SNR, although practical measurements often reveal more complex dependencies. However, through theoretical analysis, this paper will demonstrate that evaluating *SNR per unit acquisition time* provides a more meaningful metric for clinical MRS applications than plain SNR. Given that the total patient scan time is typically limited to at most one hour, the available total acquisition time (TA) for spectroscopic exams in our institute and other sequences is limited to a maximum of 10-15 minutes. The SNR per unit of acquisition time, which normalizes achievable SNR by the total acquisition time, is therefore the most relevant parameter for optimizing the SNR of clinical MRS measurements, given a maximum total tolerable acquisition time.

Experimental linear SNR per time unit gain with respect to $B_0$ have been reported in the past for fast MRSI with correction of relaxation effects and a time-domain matched filter [11], [12]. However, a theoretical analysis on the relationship between the *maximum obtainable SNR per time unit of acquisition time* (further denoted by $SNR_t$) with respect to $B_0$ has been lacking. A valid question is whether the signal-to-noise ratio per unit time of acquisition time ($SNR_t$) *always* increase as the magnetic field $B_0$ increases?

Therefore, this paper will systematically analyze the theoretical limitations and gains of *in vivo* MRS signals as a function of increasing magnetic field strength $B_0$, contrasting ultra-high, mid, and low field. More concretely it will examine to what extend the theoretical $SNR_t$ improvement of the MR-signal can be obtained in reality. More concretely, it will be investigated how factors like e.g., total available acquisition time (TA) or SAR-limitations influence the practical obtainable $SNR_t$ in a typical SVS sequence is. Given the fact that the local and global SAR increases with increasing $B_0$, the effect of $B_0$ is also included in the analysis. It is essential to emphasize that the intended RF pulse flip angles are perfectly achieved.

## Theory and Methods

### Johnson–Nyquist noise (thermal noise)

Thermal noise in MRS/MRI arises mainly from two components: the thermal motion of charged particles within the measured sample (e.g., patient tissue) and the receiver coil. The variance $\Psi^2$ of the induced noise voltage per Hz is given by [13], [14], [15]:

$$\Psi^2 = 4k_B \cdot (R_s T_s + R_c T_c) \tag{3}$$

Where $k_B$ is the Boltzmann constant, $R_s$ and $R_c$ are the equivalent noise resistance of the sample (e.g., living tissue) and the receiver coil, $T_s$ and $T_c$ are the thermodynamic temperatures of the sample and the receiver coil. It has been established that [15]

$$R_s \propto \omega^2 \propto B_0^2 \tag{4}$$

$$R_c \propto \sqrt{B_0} \tag{5}$$

Therefore:

$$\Psi^2 \propto \left(T_s B_0^2 + \alpha T_c B_0^{1/2}\right) \tag{6}$$

where α is the coil and preamplifier noise contribution coefficient, representing the additional noise sources beyond the sample. The relative noise contribution of the coil to the total noise is larger at low main magnetic field since it is proportional to $B_0^{1/2}$. For fields higher than 0.24T (10 MHz for proton) [1], [15], the noise contribution from the coil is negligible compared to that from the sample, allowing the second term to be ignored.

$$\Psi \propto B_0 \qquad (7)$$

## Pulse sequence dependent signal and noise levels

Different types of pulse sequences influence the observed signal intensities and noise levels. For single-voxel spectroscopy (SVS) sequences such as PRESS [16] and (semi) SADLOVE/LASER[17], [18], the factors [1]$\sigma_{seq}$ influencing the final observable SNR are given in [15]:

$$M_0\sqrt{N \cdot T_{sp}} \cdot F \propto n\sqrt{N \cdot T_{sp}} \cdot F = \rho \Delta V \sqrt{N \cdot T_{sp}} \cdot F = \sigma_{seq} \qquad (8)$$

in which $\Delta V$ is the volume of the spectroscopic voxel (assuming a homogeneous spin density $\rho$), $n$ is the number of spins, $N$ is number of excitation (sequence repetitions), $T_{sp}$ is the total sampling time, and $F$ is a pulse sequence dependent weighting function that depends on the pulse sequence parameters TR, TE, and inversion time TI, but also of the relaxation times $T_1$, and $T_2$. This weighting function $F$ directly influences the MR-signal amplitudes of the metabolites, and plays a central role in the computation of the SNR.

## $F$-factor and Ernst angle

It is important to note that this $F$-factor accounts for *saturation* effects of the spin system under investigation. For one single SVS signal acquisition applied on the equilibrium state magnetization $M_0$, the value of $F = 1$. If the SNR of one single MRS measurement is insufficient, the pulse sequence must be repeated, and the SRN must be improved by signal averaging. Let $TR$ represent the sequence repetition time of the sequence, and $T_1$ the relaxation time of the spin system under investigation. When $TR$ is less than approximately five times the value of $T_1$, the spin system does not fully relax to the ground state, resulting in an equilibrium longitudinal magnetization $M_z < M_0$ and, consequently, $F < 1$. In such cases, the optimal SNR cannot be achieved with a 90-degree flip angle. Instead, a smaller flip angle is required,

---

[1]Note that $\sigma_{seq}$ is the SNR per square root of frequency ($1/\sqrt{Hz}$)

known as the Ernst angle $\theta_E$ (named after Nobel Laureate Richard Ernst), maximizes the SNR under these conditions.

For an SVS sequence without selective inversion (i.e., a non-saturation inversion MRS experiment) and ignoring $T_2$ relaxation effects, an excitation pulse with a flip angle $\theta$ equal to the Ernst angle $\theta_E$ results in a *maximum* F-factor given by:

$$F = \frac{1 - e^{-TR/T_1}}{1 - \cos\theta_E \cdot e^{-TR/T_1}} \sin\theta_E \tag{9}$$

$$\theta_E = \arccos(e^{-TR/T_1}) \tag{10}$$

## SNR

Combining equations (7) and (8), the SNR of an SVS acquisition can be expressed as:

$$\text{SNR} = \frac{S_{NMR} \cdot \sigma_{seq}}{\Psi} \propto B_0 \cdot \sigma_{seq} = B_0 \cdot \rho \, \Delta V \cdot \sqrt{N \cdot T_{sp}} \cdot F \tag{11}$$

Where $S_{NMR}$ is the obtainable NMR signal from the perspective of physics and hardware, and are related to the main magnetic field strength [15], the variance $\Psi^2$ of the induced noise voltage squared per Hz is described in references [13], [14], [15], and [2]$\sigma_{seq}$ is the factor which is dependent on the MRS sequence parameters that determine the observable signal intensities and noise levels of the hardware and subject.

## SNR per unit time of acquisition time assuming a fixed TA

From a clinical perspective, the total acquisition time (TA) for a sequence is of great importance. This is because the maximum total scan duration for a patient in an MR scanner should be at all means tolerable, is approximately one hour, during which all necessary sequences must be completed to address the clinical questions. For a sequence to be considered clinically relevant, its acquisition time (TA) should ideally not exceed 10 minutes. This implies that TA for clinical sequences is essentially fixed. Consequently, it becomes crucial to evaluate the SNR per unit time of acquisition time $SNR_t$ under the constraint of a fixed TA and assess the effective SNR gain or loss as a function of magnetic field strength $B_0$. The total acquisition time TA is expressed as (with $N$ being the number of sequence repetitions):

$$TA = N \cdot TR \tag{12}$$

---

[2] Note that $\sigma_{seq}$ is expressed as the SNR per square root of frequency ($1/\sqrt{Hz}$) units.

For an SVS sequence, the SNR per unit time while keeping TA fixed (denoted as $SNR_t|_{TA}$) is given by:

$$SNR_t|_{TA} = \frac{SNR}{TA} \propto \frac{B_0\sqrt{N} \cdot F}{TA} = \frac{B_0\sqrt{TA/TR} \cdot F}{TA} \propto \frac{B_0 \cdot F(B_0)}{\sqrt{TR}} \tag{13}$$

## Specific Absorption Rate: SAR

The Specific Absorption Rate (SAR) measures the rate at which radiofrequency (RF) energy is absorbed by human tissue during an MRI or MRS(I) examination. It is a crucial safety parameter, (IEC 60601-2-33 [19]), as excessive RF exposure can lead to tissue heating and potential physiological effects. To ensure patient safety, regulatory limits on SAR restrict the RF power that can be applied, which in turn affects the choice of pulse sequences and overall acquisition efficiency. The specific absorption rate (SAR) quantifies the RF power absorbed per unit mass of human tissue, with SI units of [J.kg$^{-1}$sec$^{-1}$]. A simplified 2D model (illustrated in **Figure 1**) describes SAR as follows [20], [21]

$$\text{SAR} = \frac{\sigma A^2 \omega^2 \cdot \int_0^{T_p} B_1^+(t)^2 dt}{2m \cdot \text{TR}} = \frac{E}{2m \cdot \text{TR}} \tag{14}$$

Where $\sigma$ is the electrical conductivity of tissue, $A$ is the area of conductive loop, $\omega$ is the RF frequency (assumed constant), $B_1^+(t)$ is the RF amplitude, $m$ is the tissue mass, $T_p$ is the pulse duration, TR is the repetition time, $E$ is the RF pulse energy. Note that Eq. (14) is valid under the assumption that for a piecewise continuous RF pulse, changes in amplitude during the RF pulse are negligible compared to its carrier frequency: $|dB_1^+(t)|/dt \ll |\omega \cdot B_1^+(t)|$.

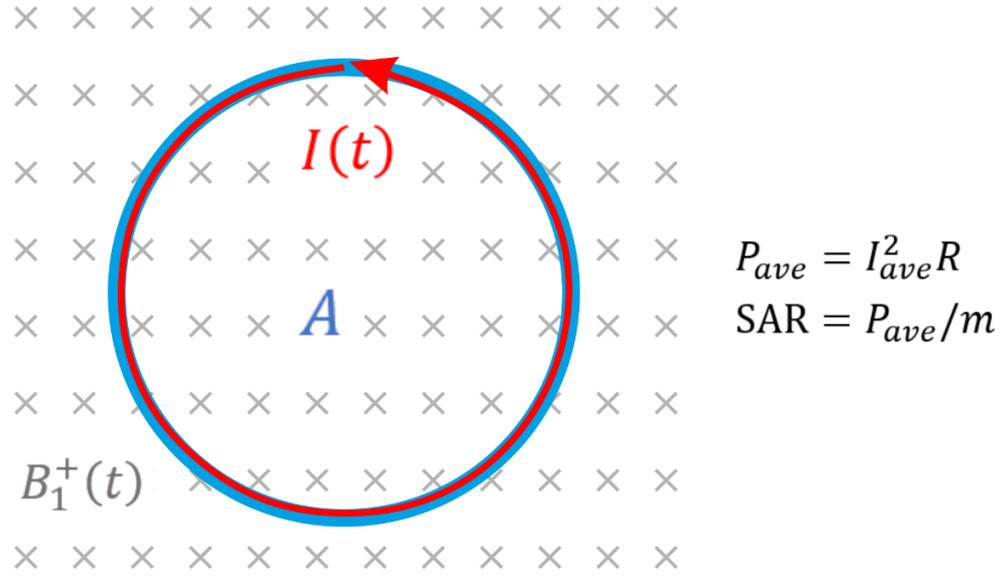

**Figure 1**: The SAR of a simplified 2D model that neglects effects such as capacitive effects and displacement currents. The conductive loop in the sample is marked in blue, with an area of A and a resistance $R$. The induced current is marked in red. This 2D model is inspired by [20].

Although the calculation of SAR in the human body is more complex, the equation above is sufficient to illustrate and investigate the relevant factors that determine how SAR limitations affect the maximum obtainable $SNR_t$. When applying MR measurement on humans, it is important to note that it is not possible to change the value of $\sigma, A, m$. Hence, to reduce SAR, one can only reduce $\omega$ or $B_1^+$, or increase TR.

$$\text{SAR} \propto \frac{\omega^2 \cdot \int_0^{T_p} B_1^+(t)^2 dt}{\text{TR}} \qquad (15)$$

For a given nominal flip angle and pulse shape, the RF pulse bandwidth is proportional to the above integral term:

$$\Delta\omega_{RF} \propto \int_0^{T_p} B_1^+(t)^2 dt \tag{16}$$

### Scaled RF pulse with magnetic field strength

Since the Larmor frequencies $\omega$ for all spin systems increase proportional to $B_0$, the pulse bandwidth $\Delta\omega_{RF}$ must be scaled by the same $B_0$-scaling factor to maintain consistent chemical shift displacement artifacts (CSDA). Additionally, for adiabatic pulses, it is assumed that their adiabaticity [22], [23], [24] should remain unchanged across different magnetic field strengths. Pulses that satisfy both criteria are referred to as *scaled RF pulses*.

$$\Delta\omega_{RF} \propto B_0 \tag{17}$$

Therefore:

$$\text{SAR} \propto \frac{\omega^2 \cdot \int_0^{T_p} B_1^+(t)^2 dt}{TR} = \frac{\omega^2 \cdot \Delta\omega_{RF}}{TR} \propto \frac{B_0^3}{TR} \tag{18}$$

### Results

In order to systematically analyze relationship between $SNR_t$ and $B_0$, this paper will analyze three different experimental boundary conditions: (**I.**) fixed TA, constant $T_1$ and constant $B_0$; (**II.**) maximum SAR, fixed TA, $B_0$ scaled-RF pulse bandwidth $\Delta\omega_{RF}$, constant $T_1$, but varying $B_0$; and (**III.**) maximum SAR, fixed TA, $B_0$ scaled-RF pulse bandwidth $\Delta\omega_{RF}$, $B_0$ dependent $T_1$, but varying $B_0$.

#### Condition I: fixed TA, constant $T_1$ and constant $B_0$

In this scenario, the total acquisition time (TA) is fixed while both the longitudinal relaxation time ($T_1$) and the magnetic field strength ($B_0$) remain constant. Under these conditions, the normal saturation effect is observed: repeated excitations lead to incomplete recovery of the equilibrium magnetisation, resulting in a progressive reduction in signal intensity. Consequently, the $SNR_t$ decreases as the repetition time (TR) increases.

Figure 2 illustrates the signal intensity within one TR (left panel) and the corresponding SNR per unit time as a function of TR (right panel) under these fixed conditions. Specifically, with these constant parameters, Eq. (13) simplifies to:

$$SNR_t|_{TA} \propto \frac{B_0}{\sqrt{TR}} \tag{19}$$

This relationship arises because the saturation of the spin system causes the available magnetization—and hence the SNR—to scale inversely with the square root of TR. Under these conditions, the $F$-factor is given by:

$$F \simeq 1, (TR > 5T_1) \tag{20}$$

These equations capture the expected normal saturation effect, where the limited recovery of magnetization during shorter TRs results in a decrease in SNR per unit time, as clearly depicted in Figure 2.

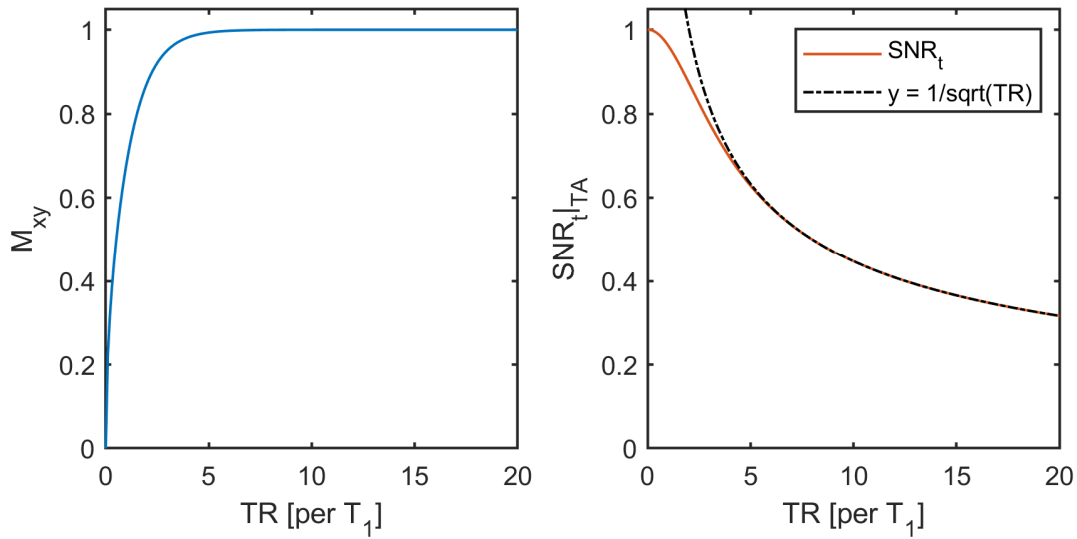

**Figure 2**: Simulations of expected transverse magnetization and $SNR_t$ assuming (i.) fixed total acquisition time *TA*, (ii.) a constant $T_1$ and (iii.) a constant $B_0$: **Left)** Signal of a SVS sequence within 1 *TR* with Ernst angle excitation. **Right)** $SNR_t$ as a function of *TR*: the total acquisition time (*TA*) is fixed so that the shorter the *TR*, the more repetitions of the excitation (number of repetitions = *TA/TR*).

**Condition II: maximum SAR, fixed *TA*, $B_0$ scaled-RF pulse bandwidth $\Delta\omega_{RF}$, constant $T_1$, but varying $B_0$**

Here we investigate the effect of increasing the $B_0$ on the $SNR_t$ where we *additionally* assume that the pulse sequence is applied at *maximum tolerable* SAR but with scaled the RF-pulse bandwidths $\Delta\omega_{RF}$ proportional to $B_0$ to maintain identical spectral coverage $\Delta ppm$. Furthermore, here we assume that the $T_1$ does not change as a function of $B_0$ as well as keeping the total acquisition time *TA* constant at all examined $B_0$. Finally, all excitations are caried out at the respective Ernst flip angles.

As concluded from the analysis of *'Condition 1'* above, $SNR_t \mid_{TA}$ always decreases with increasing *TR*. Therefore, the optimal $SNR_t \mid_{TA,opt}$ for any given SVS sequence is achieved at the shortest possible *TR*. However, in practice, this minimum *TR* is constrained by the maximum tolerable SAR. Under these conditions, the optimal $SNR_t \mid_{TA,opt}$ can be again expressed as (the same with Eq. (13)):

$$SNR_t \mid_{TA,opt} \propto \frac{B_0 \cdot F(B_0)}{\sqrt{TR}} \qquad (21)$$

and according to Eq.(18) the maximum tolerable $SAR_{max}$ is proportional to:

$$SAR_{max} \propto \frac{B_0^3}{TR} \qquad (22)$$

In Eq. (22), the maximum tolerable SAR ($SAR_{max}$), as defined by IEC 60601-2-33 [19], is considered constant under the assumption of a fixed organ and a healthy subject. By substitution of Eq. (22) into Eq. (21), the following expression for $SNR_t \mid_{TA,opt,maxSar}$ is obtained:

$$SNR_t \mid_{TA,opt,maxSar} \propto \frac{B_0 \cdot F(B_0)}{\sqrt{B_0^3/SAR_{max}}} \propto \frac{F(B_0)}{\sqrt{B_0}} \qquad (23)$$

Please note, again, that in this expression $F$ is a function of $B_0$. For a given SVS sequence with maximum SAR at a magnetic field strength $B_0$. Substitution of the $F$-factor given by Eq. (9); we find the following expression for $SNR_t \vert_{TA,opt,maxSar}$:

$$SNR_t \vert_{TA,opt,maxSar} = S_0 \cdot \frac{F(B_0)}{\sqrt{B_0}}$$
$$= S_0 \cdot \frac{1}{\sqrt{B_0}} \cdot \frac{1 - e^{-TR/T_1}}{1 - \cos(\arccos(e^{-TR/T_1})) \cdot e^{-TR/T_1}} \sin\left(\arccos(e^{-TR/T_1})\right) \quad (24)$$

In Eq. (24), $S_0$ is the initial voltage measured by the receiver system, and it is a constant which does not depends on sequence parameters and magnetic field. If the magnetic field $B_0$ is increased by a factor $f$:

$$B_0(f) = f \cdot B_{0,0} \quad (25)$$

Where $B_{0,0}$ represents the *reference magnetic field strength*, defined as the lowest $B_0$ value in the analysis ($f = 1$) In addition, we refer *the shortest TR at maximum SAR* for a pulse sequence at $B_{0,0}$ as *reference TR* (denoted as $TR_0$). According to Eq. (22):

$$TR_e = TR_0 \cdot \left(\frac{B_0}{B_{0,0}}\right)^3 = TR_0 \cdot f^3 \quad (26)$$

Where $TR_e$ is the *effective TR* which is *the shortest TR at maximum SAR* for a pulse sequence at $B_0$. By substitution of Eq.(25-26) into Eq.(24):

$$SNR_t \mid_{TA,opt,maxSar}$$

$$= S_0 \cdot \frac{1}{\sqrt{f \cdot B_{0,0}}} \qquad (27)$$

$$\cdot \frac{1 - e^{-f^3 \cdot TR_0/T_1}}{1 - \cos\left(\arccos(e^{-f^3 \cdot TR_0/T_1})\right) \cdot e^{-f^3 \cdot TR_0/T_1}} \sin\left(\arccos(e^{-f^3 \cdot TR_0/T_1})\right)$$

**Figure 3** shows the $SNR_t \mid_{TA,opt,maxSar}$ changes as a function of $f$ with different reference values $TR_0$. The $SNR_t \mid_{TA,opt,maxSar}$ is normalized to 1 at $f = 1$. Given that the maximum SAR is constant for all $B_0$, the shorter the $TR_0$, the lower the absorbed RF energy per TR is. The lower the absorbed RF energy per TR of the sequence, the greater the $SNR_t$, as can be seen **Figure 3** when the magnetic field strength $B_0$ is increased. From this figure it is evident that in the case of long $TR_0$ ($TR_0 \geq 3T_1$), there is no further gain in $SNR_t$ to be expected in case the magnetic field strength is increased. The reason for this is that, in this regime, the $SNR_t \mid_{TA,opt,maxSar}$ decreases with the square root of $f$ (Eq.(28)):

$$SNR_t \mid_{TA,opt,maxSar}(f) \simeq S_0 \cdot \frac{1}{\sqrt{f \cdot B_{0,0}}} \qquad (28)$$

The maximum SNR per time unit $SNR_t \mid_{TA,opt,maxSar}$ is reached at $f_{opt}$, which corresponds to:

$$B_{0,opt} = f_{opt} \cdot B_{0,0} \qquad (29)$$

$$TR_{e,opt} = f_{opt}^3 \cdot TR_0 \qquad (30)$$

Where $TR_{e,opt}$ is the *optimal effective TR* (in short: optimal TR) at optimal magnetic field strength $B_{0,opt}$.

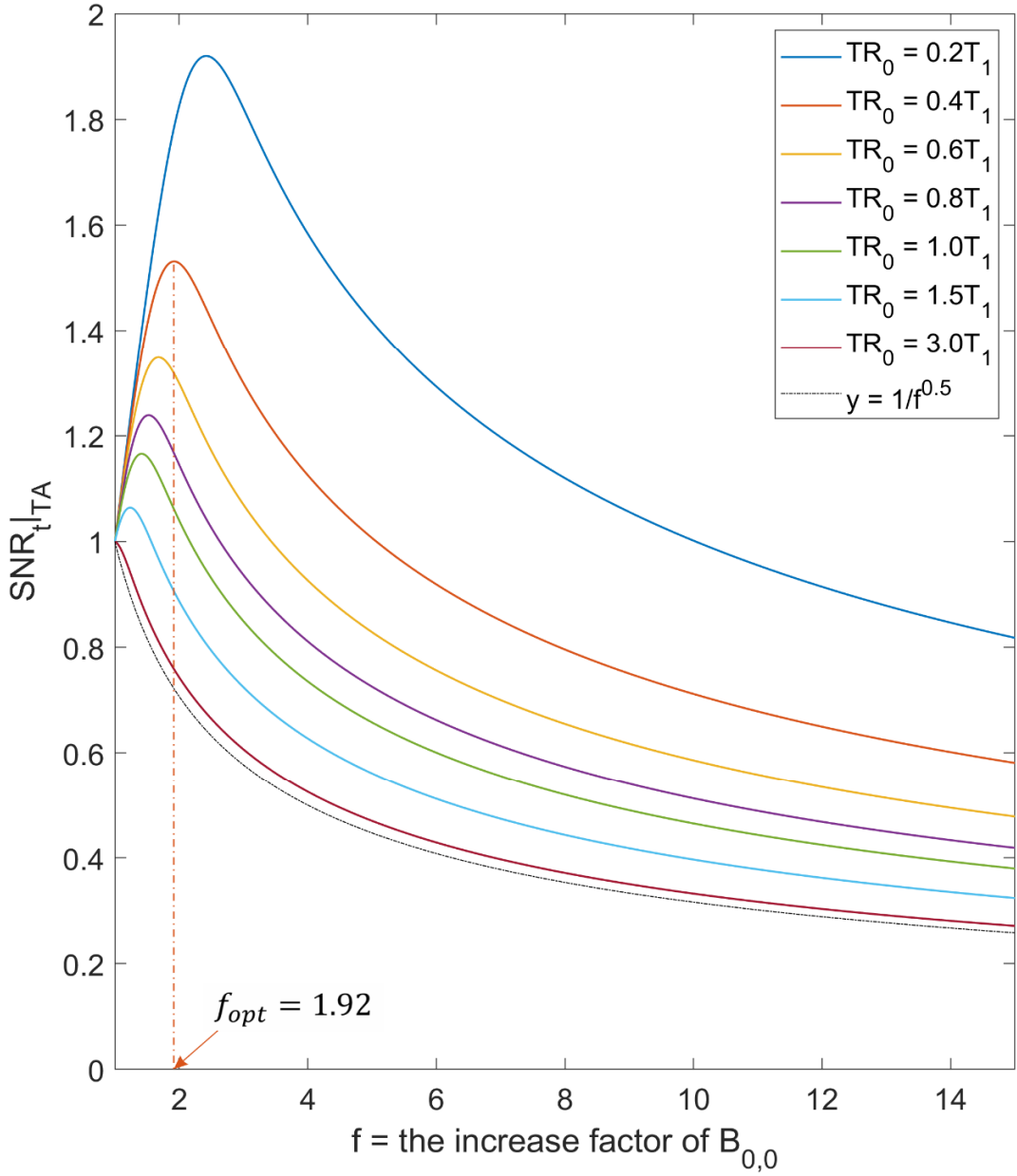

**Figure 3**: The optimal SNR per time unit along magnetic field $B_0$ with constant $T_1$. The optimal SNR per time unit of different reference $TR_0$ is normalized at $B_{0,0}$. Please note the total acquisition time $TA$ is assumed to be constant in all computations and are valid for recording spectra at maximum tolerable SAR. For instance, in the case of $TR_0 = 0.4T_1$, the optimal value $f_{opt} \simeq 1.92$, which corresponds the optimal effective $TR_{e,opt} = f_{opt}^3 \cdot TR_0 = f_{opt}^3 \cdot 0.4T_1 \simeq 2.85T_1$.

**Condition III: maximum SAR, fixed TA, $B_0$ scaled-RF pulse bandwidth $\Delta\omega_{RF}$, $B_0$ dependent $T_1$, but varying $B_0$**

We will now investigate the dependency of $SNR_t$ dependence on $T_1$, i.e. $T_1 = T_1(B_0)$. As far as we know, there is no established analytical equation for the relationship between longitudinal relaxation time $T_1$ *in vivo* and magnetic field $B_0$. Therefore, we will make some assumptions of the $B_0$ dependence of the longitudinal relaxation time $T_1$ on $B_0$:

$$T_1(f) = T_{1,0} \cdot f^\alpha \tag{31}$$

In Eq. (31), we assume $\alpha > 0$. Again, $f$ is the increase factor by which the magnetic field $B_0$ is multiplied, $T_{1,0}$ is the reference relaxation time, [3]$\alpha$ is a constant parameter.

In order to illustrate the effect of $B_0$ dependent $T_1$ on the $SNR_t|_{TA,opt,maxSar}$, we performed two series of simulations. In the first series, we assume that $T_1 \propto \sqrt{B_0}$; in the second series, we assume that $T_1 \propto B_0$. For each assumption, we tried to determine that $TR_{e,opt}$ which maximizes the $SNR_t|_{TA,opt}$. **Figure 4** shows the $SNR_t|_{TA,opt}$ with $\alpha$ equal to 0.5 and 1, respectively. As $\alpha$ increases, $T_1$ also increases with rising $B_0$, resulting in a reduced SNR gain per unit time. The underlying reasons for this behavior will be elaborated in the subsequent discussion.

We further investigated the effect of increasing $\alpha$ with a reference $TR_0 = 0.5T_{1,0}$, in order to evaluate the general effect of $T_1(\alpha)$ on the $SNR_t|_{TA,opt,maxSar}$. **Figure 5** shows the SNR per unit time $SNR_t|_{TA,opt,maxSar}$ simulation results which were performed with a reference $TR_0 = 0.5T_{1,0}$, and variable $\alpha \in [0.1, 0.7]$. A larger α reduces the maximum SNR per unit time $SNR_t|_{TA,opt,maxSar}$, but also increases the optimal field strength, $B_{0,opt} \propto f_{opt}$, at which this maximum is reached.

---

[3] Some experimental studies suggest the $\alpha$ is around 1/3, but it varies from metabolite to metabolite.

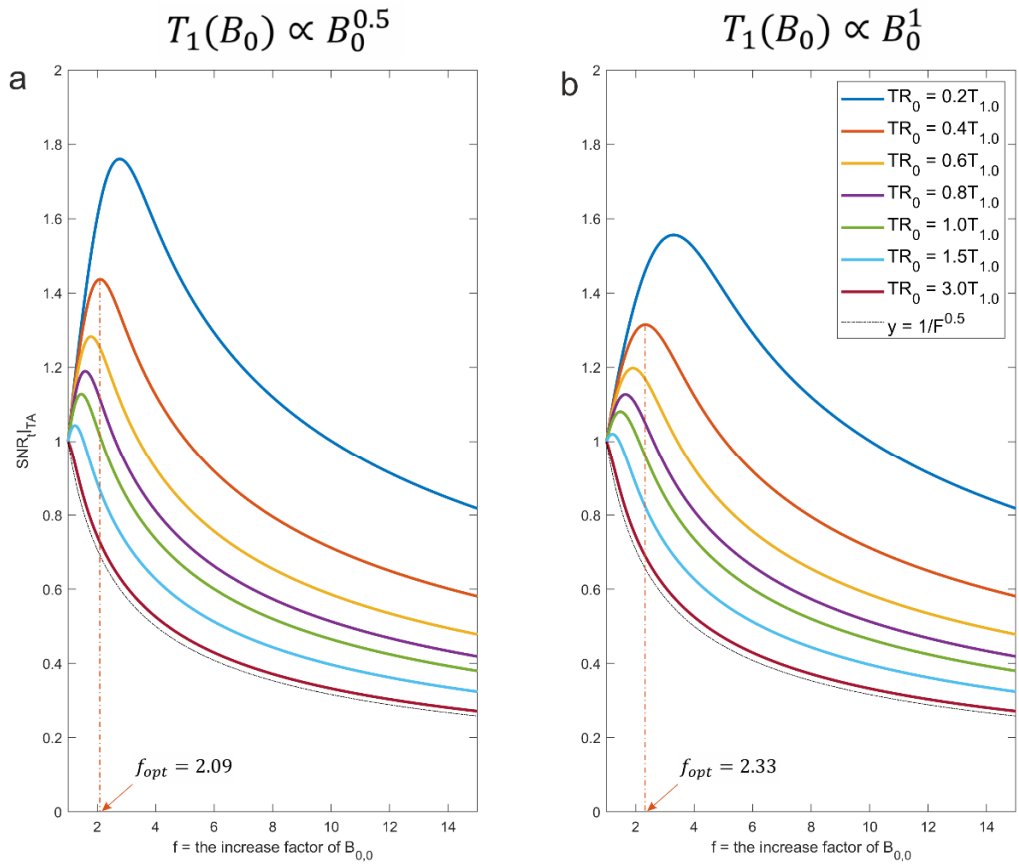

**Figure 4**: The optimal SNR per time unit along magnetic field $B_0$ with increasing $T_1$ assuming that the total acquisition time TA is constant and is performed at maximum tolerable SAR. **a)** $T_1$ is proportional to $B_0^{0.5}$. **b)** $T_1$ is proportional to $B_0$. For example, **(I)** in the case of $T_1(B_0) \propto B_0^{0.5}$ and $TR_0 = 0.4T_{1,0}$ (orange line), the $f_{opt}$ is 2.09, which corresponds the optimal effective $TR_{e,opt} = f_{opt}^3 \cdot TR_0 = f_{opt}^3 \cdot 0.4T_{1,0} \simeq 3.67T_{1,0} \simeq 2.54T_1$; **(II)** in the case that $T_1(B_0) \propto B_0^1$ and $TR_0 = 0.4T_{1,0}$ (orange line), the $f_{opt}$ is 2.33, which corresponds the optimal effective $TR_{e,opt} = f_{opt}^3 \cdot TR_0 = f_{opt}^3 \cdot 0.4T_{1,0} \simeq 5.07T_{1,0} \simeq 2.17T_1$.

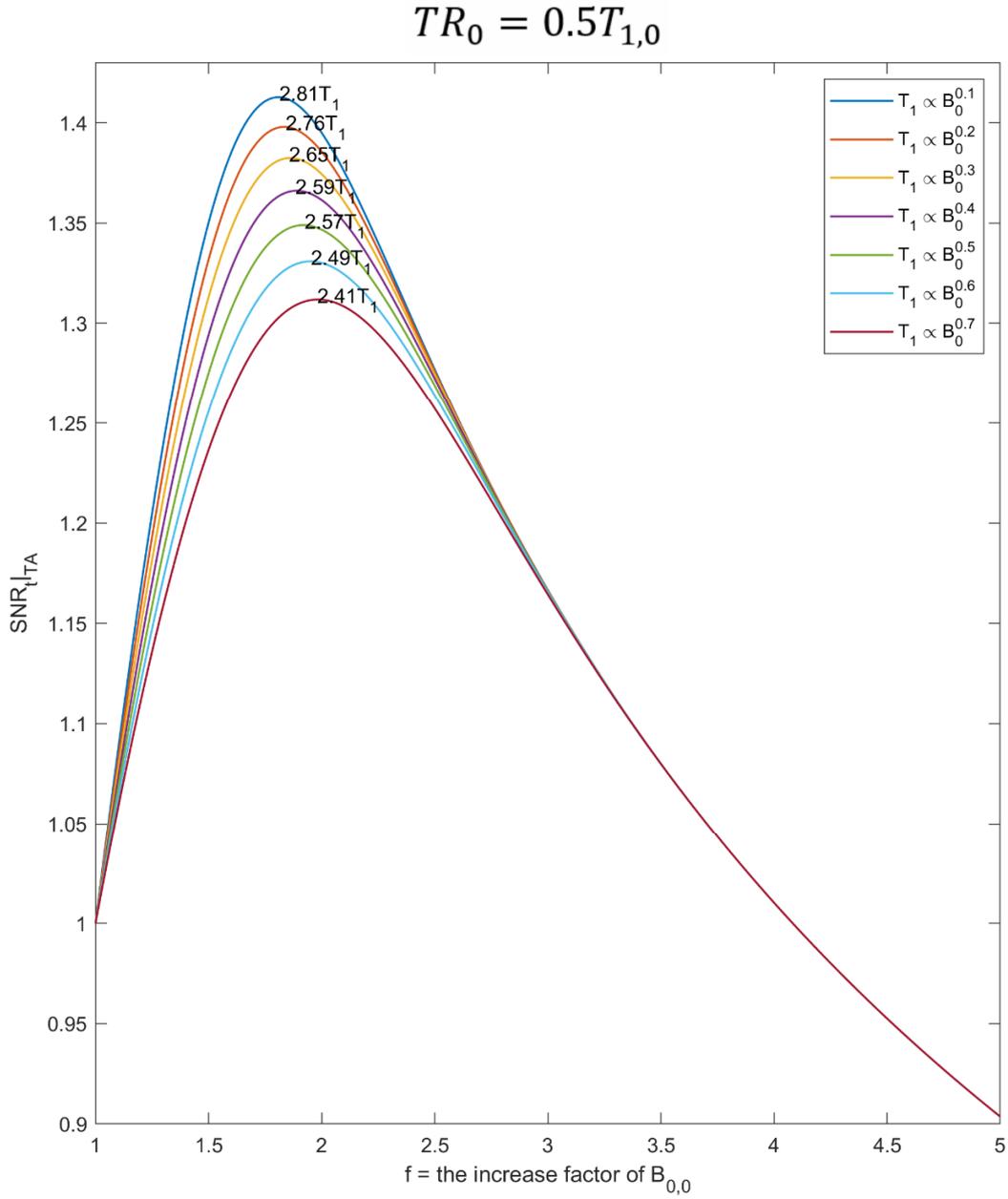

**Figure 5**: The optimal $SNR_t \mid_{TA,opt,maxSar}$ as a function of variable $\alpha$-values. Please note the total acquisition time TA is assumed is constant and is applied at maximum tolerable SAR. The optimal effective $TR_{e,opt}$ at optimal magnetic field are 2.81, 2.76, 2.65, 2.59, 2.57, 2.49 and 2.41$T_1$ for α = 0.1, 0.2, 0.3, 0.4, 0.5, 0.6, and 0.7, respectively.

To illustrate the influence of α on the maximum SNR per unit time $SNR_t|_{TA,opt,maxSar}$, we selected α=0.3 and 0.5 as numerical examples, along with the corresponding optimal scaling factor $f_{opt}$ and optimal effective repetition time $TR$ ($TR_{e,opt}$), as summarized in **Table 1**. For instance, consider an SVS sequence with following parameters at a reference magnetic field strength of $B_{0,0} = 1.5T$: $T_{1,0} = 1$ second, $TR_0 = 0.2T_{1,0} = 0.2$ seconds. In the scenario where α = 0.3, the pulse bandwidth-scaled version of this sequence achieve its maximum SNR per unit time $SNR_t|_{TA,opt,maxSar}$ at $f_{opt} = 2.61$, corresponding to a field strength of $B_0 = f_{opt} \cdot B_{0,0} = 3.915T$. The optimal effective TR is $TR_{e,opt} = 17.8 \cdot 0.2 = 3.56$ seconds, which is 17.8 times longer than $TR_0$. At this optimal field strength, the $SNR_t|_{TA,opt,maxSar}$ is increased by a factor of 1.83 compared to its value at 1.5T.

**Table 1**: The values of the maximum SNR per time unit $SNR_t|_{TA,opt,maxSar}$, and the corresponding optimal *f* and optimal effective TR for two different α values. Please note again that the values given are valid assuming a constant total acquisition time TA performed at maximum SAR.

|  | α = 0.3 | | | α = 0.5 | | |
| --- | --- | --- | --- | --- | --- | --- |
| $TR_0[T_{1,0}]$ | $f_{opt}$ | $TR_{e,opt}[TR_0]$ | $SNR_t|_{TA,opt,maxSar}$ | $f_{opt}$ | $TR_{e,opt}[TR_0]$ | $SNR_t|_{TA,opt,maxSar}$ |
| 0.2 | 2.61 | 17.8 | 1.83 | 2.77 | 21.3 | 1.76 |
| 0.4 | 2.02 | 8.22 | 1.48 | 2.10 | 9.24 | 1.44 |
| 0.6 | 1.74 | 5.25 | 1.31 | 1.79 | 5.69 | 1.28 |
| 0.8 | 1.56 | 3.80 | 1.21 | 1.59 | 4.04 | 1.19 |
| 1.0 | 1.44 | 2.99 | 1.14 | 1.46 | 3.09 | 1.13 |
| 1.5 | 1.24 | 1.91 | 1.05 | 1.24 | 1.91 | 1.04 |
| 2.0 | 1.11 | 1.38 | 1.01 | 1.10 | 1.35 | 1.01 |

## Discussion

Condition 2 — We first discuss the $SNR_t|_{TA,opt,maxSar}$ results of condition 2, i.e. the spectroscopy pulse sequence is applied at fixed maximum SAR, fixed TA, constant $T_1$, $B_0$-scaled RF pulse bandwidth $\Delta\omega_{RF}$, and varying $B_0$. Under these conditions the $SNR_t|_{TA,opt,maxSar}$ shows a single maximum as a function of $B_0$ as shown in **Figure 3**. This means that:

$$\frac{dSNR_t|_{TA,opt,maxSar}}{dB_0} = \frac{d(F(B_0)/\sqrt{B_0})}{dB_0} = \begin{cases} > 0, & B_0 < B_{0,opt} \\ = 0, & B_0 = B_{0,opt} \\ < 0, & B_0 > B_{0,opt} \end{cases} \quad (32)$$

Function analysis reveals that for a sequence with a short reference $TR_0$ (and $TR_0$ is the shortest possible TR, limited by SAR-limitation, at current reference magnetic field $B_{0,0}$), this is because the factor $F(B_0)$ increases faster than $\sqrt{B_0}$ up to the optimal magnetic field $B_{0,opt}$ (Eq.(23)):

For $B_0 < B_{0,opt}$ the increase of SNR per time unit is primarily driven by the SNR gain due to the rising $B_0$. When $B_0 > B_{0,opt}$ the saturation of the spin system starts to counteract and eventually dominates the $B_0$ related SNR increase, leading to a net decrease in SNR per time unit of measurement time.

The optimal effective TR ($TR_e = f^3 \cdot TR_0$) for a sequence at the optimal magnetic field $B_{0,opt}$ with a constant $T_1$ typically falls around $2.8T_1 - 3T_1$ (see **Figure 3**). Thus, for sequences with the shortest possible TR, but still greater than $3T_1$, increasing the magnetic field $B_0$ further yields no additional gain in SNR per time unit.

Condition 3 — To recall, in this case we additionally assume that the $T_1$ of the metabolites are $B_0$ dependent as well, i.e. $T_1 = T_1(B_0)$. For a given $TR_0$, *and* assuming that $T_1$ rises with increasing $B_0$, and the optimal effective TR ($TR_{e,opt}$) for maximum SNR per time unit increases (**Figure 4**). In the case of $TR_0 = 0.4T_{1,0}$, the optimal $TR_{e,opt}$ values are $3.67T_{1,0}$ and $5.07T_{1,0}$ (note $T_{1,0}$ is constant) for $T_1 \propto B_0^{0.5}, B_0^1$, respectively. That means, as $\alpha$ increases, so does the optimal TR ($TR_{e,opt}$). On the other hand, $T_1$ also increases with $B_0$ when $\alpha > 0$, but faster than the optimal TR (Eq.(33)). Consequently, the ratio of $TR_{e,opt}$ to $T_1$ decreases with increasing $\alpha$ (**Figure 5**).

$$\frac{d\left(T_1(B_0)/TR_{e,opt}(B_0)\right)}{dB_0} > 0 \quad (33)$$

In clinical practice, many SVS sequences have a TR of approximately 3 seconds or more at 3T [25], which is longer than $2T_1$ at this field strength. Therefore, further gains in SNR per time unit ($SNR_t$) are not expected when scaling such sequences to higher magnetic field strengths, as indicated in **Table 1**. Specifically, for a sequence which TR $= 2T_1$, the additional gain in $SNR_t$ from increasing $B_0$ is only 1% assuming $T_1 \propto B_0^\alpha$ with $\alpha = [0.3, 0.5]$. The reason for this is that the bandwidth of RF pulses also scales with $B_0$ causing the effective TR to increase with $B_0^3$.

Practical case examples — To illustrate simulated results of the theoretical model, we consider a practical example using an SVS sequence at a reference magnetic field ($B_{0,0} = 1.5$T) with the following parameters: acquisition time (*TA*) of 5 minutes, reference relaxation time ($T_{1,0}$) of 1 second, reference repetition time ($TR_0$) of 0.4 seconds, and 750 excitations (repetitions).

In the case of *Condition 2* (constant $T_1$, as shown in **Figure 3**), the dependence of $SNR_t$ on $B_0$ is represented by the orange line ($TR_0 = 0.4T_1$). The optimal signal-to-noise ratio per unit time ($SNR_t|_{TA,opt,maxSar}$) is achieved at an optimal magnetic field of $B_{0,opt} = 2.9$T, with an optimal effective TR ($TR_{e,opt}$) of 2.9 seconds, and ~103 excitations (repetitions). At this optimal field strength (2.9T), $SNR_t|_{TA,opt,maxSar}$ increases by a factor of approximately 1.53 compared to the reference field ($B_{0,0}$) of 1.5T.

Similarly, in the case of *Condition 3* ($T_1 \propto B_0^{0.5}$, as shown in **Figure 4a**), the relationship between $SNR_t$ and $B_0$ follows the same orange line ($TR_0 = 0.4T_{1,0}$). The optimal $SNR_t|_{TA,opt,maxSar}$ occurs at $B_{0,opt} = 3.1$T, with an optimal effective TR of $TR_{e,opt} = 3.7$ seconds, and ~81 excitations (repetitions). In this scenario, $SNR_t|_{TA,opt,maxSar}$ increases by a factor of approximately 1.44 at the optimal field strength (3.1T) compared to the reference field (1.5T).

Compensation — However, as mentioned above, $B_1^+$-insensitive SVS sequences, especially those which use *slice selective* adiabatic RF pulses (like semi-SADLOVE/LASER [17], [18]) are associated with high SAR costs and already require a relatively long TR (usually larger than $1.5T_1$) at 3T. Therefore, the additional RF power of adiabatic RF pulses at higher $B_0$ must always be compensated when increasing the main magnetic field $B_0$. This can either be realized by reducing the RF-bandwidth, or reducing the RF-amplitude (but still above the minimal adiabatic $B_1^+$-threshold), or both [26]. However, reducing RF bandwidth increases chemical shift displacement artifacts (CSDA), while lowering the RF amplitude (a) for adiabatic pulses enhances their sensitivity to $B_1^+$ inhomogeneity and (b) for non-adiabatic pulses leads to reduced actual flip angles. Therefore, whether the $SNR_t|_{TA,opt,maxSar}$ gain outweighs the performance loss from RF pulse compensation remains an open question.

## Limitations

### T$_2$ decay

The $T_2$ value is also depended on magnetic field strength. Generally, higher magnetic fields result in shorter $T_2$ values, further reducing the achievable optimal SNR per unit time ($SNR_t \mid_{TA,opt,maxSar}$) when the magnetic field increases. However, the echo time (*TE*) used in SVS can vary depending on the specific metabolites of interest. For instance, the short *TE* (<35 ms) is typically employed to maximize SNR for major metabolites (such as NAA, Cho, and Cr), whereas a relatively long *TE* (68–144 ms) is required for detecting *J*-coupled metabolites (such as GABA, 2HG, and Lac) in spectral editing techniques. Thus, the effect of $T_2$ is not included in this work, since the analysis presented here is most general, and is not targeting a particular *TE*. Instead, this study provides a theoretical upper bound for $SNR_t \mid_{TA,opt,maxSar}$ that is valid across all echo times and can be scaled for specific $T_2$/*TE* accordingly.

### CSI and other types of fast MRSI

One key assumption of this work is that the total acquisition time (*TA*) is fixed, and multiple averages are acquired. The other key assumption is application of the sequence at maximum tolerable SAR. The number of averages for SVS is inversely proportional to the repetition time (*TR*), calculated as: number of averages = *TA*/*TR*. If chemical shift imaging (CSI) [12] or other magnetic resonance spectroscopic imaging (MRSI) [27] protocols involve multiple averages, the analysis presented here would also apply to those cases. However, in most practical scenarios, these sequences are acquired with only a single average, placing them outside the scope of this analysis presented here. Additionally, in the case of fast MRSI [27], the acquired spectral bandwidth becomes a critical limiting factor at higher magnetic fields. Therefore, extending this analysis to fast MRSI protocols would require incorporating the effect of spectral bandwidth explicitly.

## Conclusion

This paper derived a theoretical model for the SNR per time unit of acquisition time to be expected as a function of increasing $B_0$ Under the following conditions: (1.) a fixed TA, (2) maximum SAR constraints, (3.) the bandwidth of RF pulse scaled with magnetic field strength, (4.) maintenance of adiabaticity when using adiabatic pulses, (5.) PRESS or (semi) SADLOVE/LASER type SVS sequences, the *SNR per unit measurement time* shows very little to no improvement, or may even decrease, if the *TR* exceeds $2T_1$ as the magnetic field strength increases.

## Acknowledgements

The funding was provided by Schweizerischer Nationalfonds zur Förderung der Wissenschaftlichen Forschung, grant numbers: SNSF-182569 and SNSF-207997.

# References

[1] D. I. Hoult and P. C. Lauterbur, "The sensitivity of the zeugmatographic experiment involving human samples," *Journal of Magnetic Resonance (1969)*, vol. 34, no. 2, pp. 425–433, May 1979, doi: 10.1016/0022-2364(79)90019-2.

[2] J. T. Vaughan *et al.*, "7T vs. 4T: RF power, homogeneity, and signal-to-noise comparison in head images," *Magn Reson Med*, vol. 46, no. 1, pp. 24–30, Jul. 2001, doi: 10.1002/MRM.1156.

[3] I. Tkáč, G. Öz, G. Adriany, K. Uğurbil, and R. Gruetter, "In vivo $^1$H NMR spectroscopy of the human brain at high magnetic fields: Metabolite quantification at 4T vs. 7T," *Magn Reson Med*, vol. 62, no. 4, pp. 868–879, Oct. 2009, doi: 10.1002/mrm.22086.

[4] P. B. Barker, D. O. Hearshen, and M. D. Boska, "Single-voxel proton MRS of the human brain at 1.5T and 3.0T," *Magn Reson Med*, vol. 45, no. 5, pp. 765–769, May 2001, doi: 10.1002/MRM.1104.

[5] K. Kantarci *et al.*, "Proton MR Spectroscopy in Mild Cognitive Impairment and Alzheimer Disease: Comparison of 1.5 and 3 T," *AJNR Am J Neuroradiol*, vol. 24, no. 5, p. 843, May 2003, Accessed: Aug. 02, 2022. [Online]. Available: /pmc/articles/PMC7975780/

[6] O. Gonen, S. Gruber, B. S. Y. Li, V. Mlynárik, and E. Moser, "Multivoxel 3D Proton Spectroscopy in the Brain at 1.5 Versus 3.0 T: Signal-to-Noise Ratio and Resolution Comparison," *AJNR Am J Neuroradiol*, vol. 22, no. 9, p. 1727, 2001, Accessed: Aug. 02, 2022. [Online]. Available: /pmc/articles/PMC7974443/

[7] J. Kim *et al.*, "Comparison of 1.5T and 3T 1H MR Spectroscopy for Human Brain Tumors," *Korean J Radiol*, vol. 7, no. 3, p. 156, 2006, doi: 10.3348/kjr.2006.7.3.156.

[8] S. Younis *et al.*, "Feasibility of Glutamate and GABA Detection in Pons and Thalamus at 3T and 7T by Proton Magnetic Resonance Spectroscopy," *Front Neurosci*, vol. 14, Oct. 2020, doi: 10.3389/fnins.2020.559314.

[9] S. Pradhan *et al.*, "Comparison of single voxel brain MRS AT 3T and 7T using 32-channel head coils," *Magn Reson Imaging*, vol. 33, no. 8, pp. 1013–1018, Oct. 2015, doi: 10.1016/j.mri.2015.06.003.

[10] M. C. Stephenson, "Applications of multi-nuclear magnetic resonance spectroscopy at 7T," *World J Radiol*, vol. 3, no. 4, p. 105, 2011, doi: 10.4329/wjr.v3.i4.105.

[11] R. Otazo, B. Mueller, K. Ugurbil, L. Wald, and S. Posse, "Signal-to-noise ratio and spectral linewidth improvements between 1.5 and 7 Tesla in proton echo-planar spectroscopic imaging," *Magn Reson Med*, vol. 56, no. 6, pp. 1200–1210, Dec. 2006, doi: 10.1002/MRM.21067.

[12] R. Pohmann, M. Von Kienlin, and A. Haase, "Theoretical evaluation and comparison of fast chemical shift imaging methods," *J Magn Reson*, vol. 129, no. 2, pp. 145–160, 1997, doi: 10.1006/JMRE.1997.1245.

[13] J. B. Johnson, "Thermal Agitation of Electricity in Conductors," *Physical Review*, vol. 32, no. 1, p. 97, Jul. 1928, doi: 10.1103/PhysRev.32.97.

[14]  H. Nyquist, "Thermal Agitation of Electric Charge in Conductors," *Physical Review*, vol. 32, no. 1, p. 110, Jul. 1928, doi: 10.1103/PhysRev.32.110.

[15]  W. A. Edelstein, G. H. Glover, C. J. Hardy, and R. W. Redington, "The intrinsic signal-to-noise ratio in NMR imaging," *Magn Reson Med*, vol. 3, no. 4, pp. 604–618, Aug. 1986, doi: 10.1002/MRM.1910030413.

[16]  P. A. BOTTOMLEY, "Spatial localization in NMR spectroscopy in vivo," *Ann N Y Acad Sci*, vol. 508, no. 1, pp. 333–348, 1987, doi: 10.1111/J.1749-6632.1987.TB32915.X.

[17]  J. Slotboom, A. F. Mehlkopf, and W. M. M. J. Bovée, "A single-shot localization pulse sequence suited for coils with inhomogeneous RF fields using adiabatic slice-selective RF pulses," *Journal of Magnetic Resonance (1969)*, vol. 95, no. 2, pp. 396–404, Nov. 1991, doi: 10.1016/0022-2364(91)90229-M.

[18]  M. Garwood and L. DelaBarre, "The Return of the Frequency Sweep: Designing Adiabatic Pulses for Contemporary NMR," *Journal of Magnetic Resonance*, vol. 153, no. 2, pp. 155–177, Dec. 2001, doi: 10.1006/jmre.2001.2340.

[19]  *IEC 60601-2-33*, 4.0. Geneva: International Electrotechnical Commission, 2022.

[20]  A. D. Elster, "MRIquestions.com."

[21]  T. M. Fiedler, M. E. Ladd, and A. K. Bitz, "SAR Simulations & Safety," *Neuroimage*, vol. 168, pp. 33–58, Mar. 2018, doi: 10.1016/J.NEUROIMAGE.2017.03.035.

[22]  J. Baum, R. Tycko, and A. Pines, "Broadband and adiabatic inversion of a two-level system by phase-modulated pulses," *Phys Rev A (Coll Park)*, vol. 32, no. 6, pp. 3435–3447, Dec. 1985, doi: 10.1103/PhysRevA.32.3435.

[23]  S. Conolly, G. Glover, D. Nishimura, and A. Macovski, "A reduced power selective adiabatic spin-echo pulse sequence," *Magn Reson Med*, vol. 18, no. 1, pp. 28–38, Mar. 1991, doi: 10.1002/mrm.1910180105.

[24]  P. Balchandani, J. Pauly, and D. Spielman, "Designing adiabatic radio frequency pulses using the Shinnar-Le Roux algorithm.," *Magn Reson Med*, vol. 64, no. 3, pp. 843–51, Sep. 2010, doi: 10.1002/mrm.22473.

[25]  D. Hong, S. Rohani Rankouhi, J.-W. Thielen, J. J. A. van Asten, and D. G. Norris, "A comparison of sLASER and MEGA-sLASER using simultaneous interleaved acquisition for measuring GABA in the human brain at 7T," *PLoS One*, vol. 14, no. 10, p. e0223702, Oct. 2019, doi: 10.1371/journal.pone.0223702.

[26]  G. Weng *et al.*, "SLOW: A novel spectral editing method for whole-brain MRSI at ultra high magnetic field," *Magn Reson Med*, vol. 88, no. 1, pp. 53–70, Jul. 2022, doi: 10.1002/mrm.29220.

[27]  W. Bogner, R. Otazo, and A. Henning, "Accelerated MR spectroscopic imaging—a review of current and emerging techniques," *NMR Biomed*, vol. 34, no. 5, p. e4314, May 2021, doi: 10.1002/NBM.4314.